\documentclass[secnumarabic,floatfix,nofootinbib, tightenlines,nobibnotes,aps,prl,reprint]{revtex4-1}

\usepackage{gensymb}
\usepackage{xcolor} % document color scheme
\usepackage{hyperref} % For hyperlinks in the PDF

\usepackage{graphicx}
\graphicspath{ {images_pdf/} }

\usepackage{booktabs}
\usepackage{multirow}
\usepackage{array}

\begin{document}

\title{Low-loss single-mode integrated waveguides in soda-lime glass}

\author{Dyakonov I.V.$^1$}
\email[Correspondence should be sent to: ]{iv.dyakonov@physics.msu.ru}
\author{Kalinkin A.A.$^1$}
\author{Saygin M.Y.$^{1,2}$}
\author{Abroskin A.G.$^{1}$}
\author{Radchenko I.V.$^{1}$}
\author{Straupe S.S.$^{1}$}
\author{Kulik S.P.$^{1}$}
\affiliation{$^1$Faculty of Physics, Lomonosov Moscow State University, Moscow, Russia}
\affiliation{$^2$P.N. Lebedev Physical Institute of the Russian Academy of Sciences, Moscow, Russia}

\date{\today}

\begin{abstract}
Low-loss single-mode optical waveguide fabrication process in extra-white soda-lime glass is demonstrated. Waveguiding structures are formed in bulk substrates employing femtosecond laser writing technology. The combination of a slit beam-shaping method and a multiscan fabrication process enables printing of waveguides with a well-defined symmetric cross-section profile. Fabricated waveguides exhibit 0.86 dB/cm propagation loss for 800~nm wavelength. Bending loss in the waveguides are addressed experimentally and compared with a model for radiation loss.
\end{abstract}


\maketitle

\section{Introduction \label{sec::ntro}} 

Recent progress in integrated photonic device fabrication endowed researchers from different branches of optical science with a new versatile toolbox. Optical chips found their application in telecommunication \cite{Doerr2015}, biomedical photonics \cite{Washburn2011}, quantum optics \cite{OBrien2013}, and many other areas. Shrinking complex optical systems onto a single solid-state substrate adds precision and control to the experiment, thus making the integrated platform particularly appealing for the rapidly developing field of photonic technologies.

Femtosecond laser writing (FSLW) of optical waveguides established itself as a low-cost approach for rapid prototyping of unique integrated photonic devices. The widespread availability of femtosecond lasers, both in terms of cost and experience required to handle them, makes the technique advantageous over the more demanding alternative technologies. Moreover, laser writing enables creation of complex 3D photonic structures without resorting to sophisticated multi-step processes, such as those used in lithography. Thus, the FSLW technique is considered a versatile tool.

FSLW enables machining of optical waveguiding structures in virtually any transparent material \cite{DellaValle2008}. Single-mode optical waveguide fabrication has been demonstrated in fused silica \cite{Shah2005}, borosilicate glasses \cite{Eaton2006}, phosphate glasses \cite{Chan2003}, chalcogenide glass \cite{Efimov2001}, and in a variety of doped glass substrates \cite{Schaffer2003,Taccheo2004,Psaila2007}. Soda-lime glass is an extremely low-cost material, which is commonly used, for example, in base plates for microscopy. It is usually not considered as a good material for demanding optical applications due to a large amount of contained impurities. However, specific material compound allows extra-white soda-lime glass to exhibit very high optical transmission in visible and near-infrared spectral range, thus making the fabrication of integrated optical devices by FSLW feasible as shown in \cite{Minoshima2001,Kowalevicz2005,Tong2006,Lazcano2016}.

Laser-written waveguides are formed inside the volume of a moving transparent sample by tightly focused radiation. In order for the radiation intensity to be high enough, pulsed lasers with the pulse duration typically in the picosecond and sub-picosecond range are used, for the relevant nonlinear processes \cite{Gattass2008} to leave permanent refractive index change. Several approaches to control the waveguide cross-section are known: shaping the beam and/or the wavefront profile using a cylindrical telescope \cite{Osellame2003}, a slit \cite{Ams2008} or adaptive optical devices such as a spatial light modulator (SLM) \cite{Salter2012} or a deformable mirror \cite{simmonds2011}. With such techniques a single scan along the waveguide path can be enough to form a waveguide core in the case of positive refractive index change. However, the process of refractive index change by an intense laser pulse field is an essentially non-equilibrium process that makes the modified volume inhomogeneous. Such inhomogeneities result in additional loss due to scattering and absorption. One can smooth the structure by traversing the laser spot multiple times along the predefined trajectory (\textit{multiscan} fabrication process), thereby decreasing the loss. Moreover, the cross-section profile of the waveguide can be precisely controlled by means of the multiscan inscription procedure. Multiscan waveguide writing is typically used for fabrication of depressed cladding waveguides (or type-II waveguides) in order to form a waveguide core between several tracks inscribed in the substrate \cite{Okhrimchuk2012}. The same principle has been adopted in \cite{Nasu2005,Psaila2007,Bookey2007,Brown2012} to print a waveguiding core and more complex structures such as arrayed waveguide gratings \cite{Douglass2015}.  

In this paper we report an implementation of FSLW technology to form low-loss single-mode optical waveguides in soda-lime glass with the controllable cross-section geometry. Our method employs slit beam shaping and a multiscan processing procedure to fabricate waveguides with a well-defined rectangular cross-section geometry as well as reduced bending and propagation loss. We focus on single-mode waveguides design for 800~nm wavelength, which is convenient for experiments in quantum optics, since both high-quality sources and efficient single-photon detectors (Si APDs) for this wavelength are available. We believe that the results presented here constitute a step towards a cost reduction of chip prototyping and may impact the further development of low-cost integrated photonic components.

\section{Waveguide fabrication \label{sec::waveguide_fab}}

\begin{figure*}[ht]
\includegraphics[width=\textwidth]{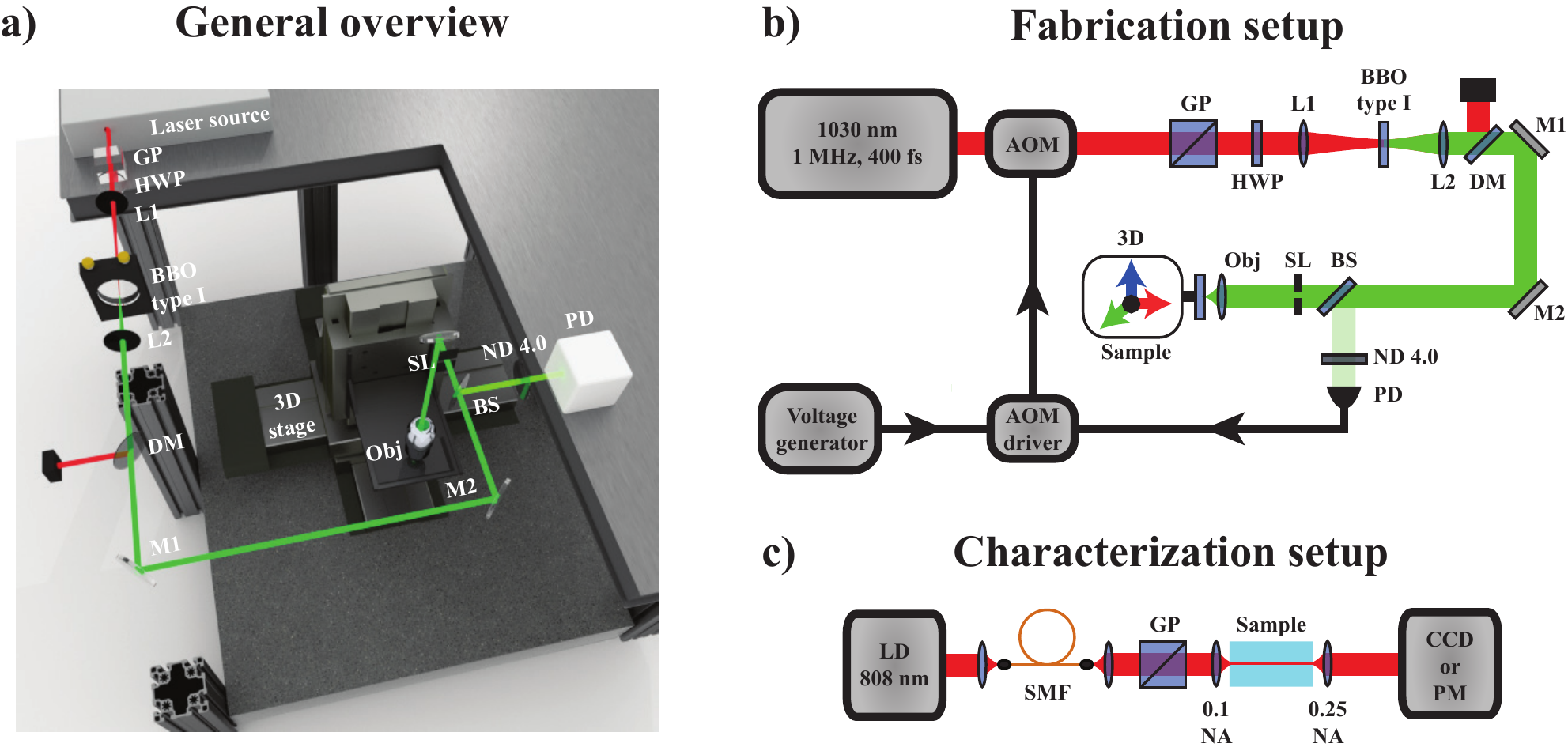}
\caption{a) General overview of the FSLW experimental arrangement. b) FSLW fabrication setup. Radiation of an ytterbium fiber laser (1030~nm, 400~fs pulse duration, 1~MHz repetition rate) is linearly polarized with a Glan prism GP, and focused by a $f=250$~mm lens L1 inside a nonlinear BBO crystal. The beam polarization is matched with the crystal fast axis by a half-wave plate HWP. Frequency-doubled laser light at 515 nm is collimated by a $f=75$~mm lens L2 and directed to a slit SL shaping the beam pattern focused by a 0.7 NA objective Obj. A beamsplitter BS reflects a small portion of light onto a photodiode PD used to monitor the current beam power. c) Waveguide characterization setup. A 808~nm laser diode beam was spatially filtered with a single-mode fiber and injected into a waveguide by a 0.1~NA microscope objective. A Glan prism sets the input light polarization state. At the output we used a 0.25~NA microscope objective to either image the end face of the chip on a CCD camera or to collimate the output light and measure its power with a power-meter PM.\label{fig::experimental_setup}}
\end{figure*}

In FSLW procedure tight laser beam focusing and short pulse length give rise to several nonlinear optical processes inside the focal spot \cite{Gattass2008}, which heavily affect the shape and the size of a modified refractive index region. Thus, single scan waveguide writing process may result in a very peculiar core cross-section geometry. For example, one may observe a large thermally-induced low-contrast zone surrounding a guiding core \cite{Eaton2008} or even multiple light-guiding regions located in different areas of a treated material sample \cite{McMillen2012}. Such specific features of waveguide cross-section can severely limit the topology of integrated optical devices fabricated with FSLW technology. Applications demanding no sophisticated optical circuitry, such as power splitters or waveguide lasers \cite{DellaValle2008}, are insensitive to cross-section artifacts if the waveguide has low loss at the desired wavelength, but some recently developed quests in the field of quantum photonics require delicate engineering of complex waveguide arrangement inside the material \cite{Crespi2013}. Developing a fabrication process of waveguides with a well-defined cross-section shape and size significantly simplifies further design of complex integrated photonic chips.

Waveguide fabrication setup used in this work is depicted in Fig.~\ref{fig::experimental_setup}. The output of an ytterbium fiber laser (Menlo Systems BlueCut, 400~fs, 1~MHz, 1030~nm) was frequency doubled with a 0.8~mm BBO crystal cut at 23.9\degree. A small portion $\approx 2\%$ of 515~nm laser light was deflected from the main optical path and directed to a photodiode PD. The photodiode tracks the time-varying output power of the laser source and provides a feedback signal for a real-time power stabilization circuit. A slit SL was placed at a distance of approximately 29 cm from the focusing objective. Laser light was focused inside the sample by an infinity corrected 0.7 NA objective with a large working distance (Mitutoyo M Plan Apo 100x). The sample was mounted on a three-axis air-bearing translation stage (AeroTech FiberGlide3D). The power of the writing beam was controlled with an acousto-optic modulator AOM integrated in the laser head. The power stabilization system allowed us to eliminate the detrimental variations in the laser power, stemmed from the instability inherent to the laser itself ($\sim 1\%$) and those caused by environmental changes during the writing process, and to ensure the stability of writing conditions over long processing times. A homebuilt external AOM driver circuit with an active feedback loop employed for active power stabilization compared PD output signal and the reference signal from a reference voltage generator and produced the AOM driving voltage. Furthermore, the presented scheme might be used for active tuning of the output power to achieve optimal writing conditions at different fabrication steps. For instance, a combination of an active slit width adjustment and a position-synchronized power modulation may substitute adaptive optic kits in 3D waveguide printing applications demanding additional writing condition correction for different processing depths inside the sample. 

Multiscan waveguide printing approach augments the standard FSLW process with an additional degree of freedom, which can be employed for cross-section profile shaping. At first, the FSLW setup has to be tuned to inscribe a uniformly modified refractive index region inside the material in a single scan process. The footprint of a single scan cross-section defines a reasonable range of transverse shifts for the sample translation trajectory in a multiscan process. Our goal was to implement a multiscan waveguide fabrication method to form waveguides with a symmetric and approximately rectangular profile.

\begin{figure}[h]
\includegraphics[width=0.8\columnwidth]{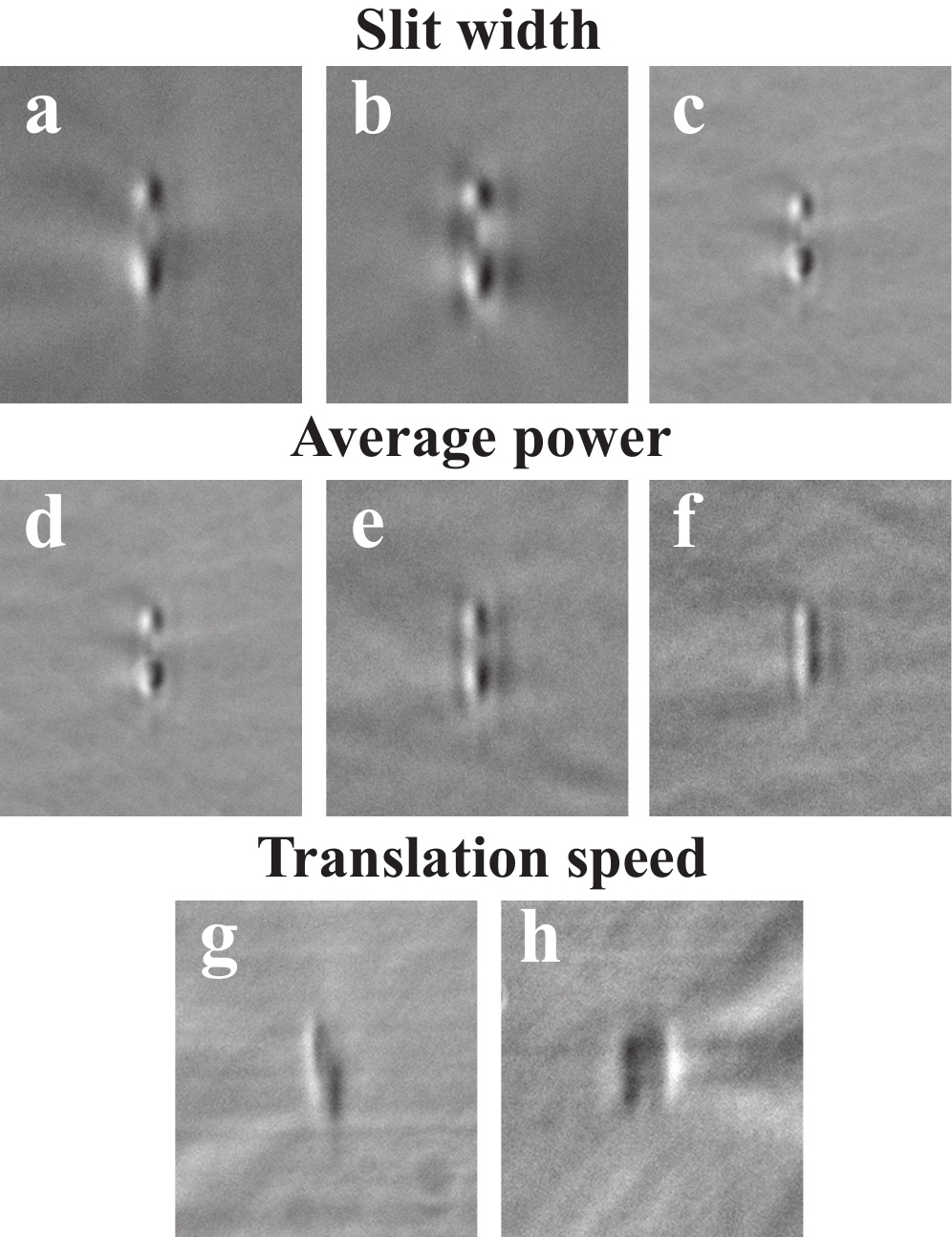}
\caption{Cross-sections images of integrated structures created with different fabrication parameters taken in an optical transmission microscope equiped with a DIC module. a)-c) illustrate cross-section shapes for different slit widths (375~$\mu$m, 350~$\mu$m and 325~$\mu$m, respectively), other exposure parameters are fixed, pulse energy of the writing beam is 52 nJ. d)-f) images show cross-section shapes of the structures fabricated after fixing the slit width at 325~$\mu$m and varying the pulse energy (results for energies of 52~nJ, 44~nJ and 34~nJ, respectively, are depicted). In a multiscan regime an effect of longitudinal shearing of the waveguide cross-section for the low translation speed (0.1 mm/s) was observed (g). For a larger speed of 1 mm/s the cross-section became more symmetric (h). Similar shearing effect was observed in \cite{Bookey2007}.\label{fig::parameter_map}}
\end{figure}

\begin{figure}[b]
\includegraphics[width=\columnwidth]{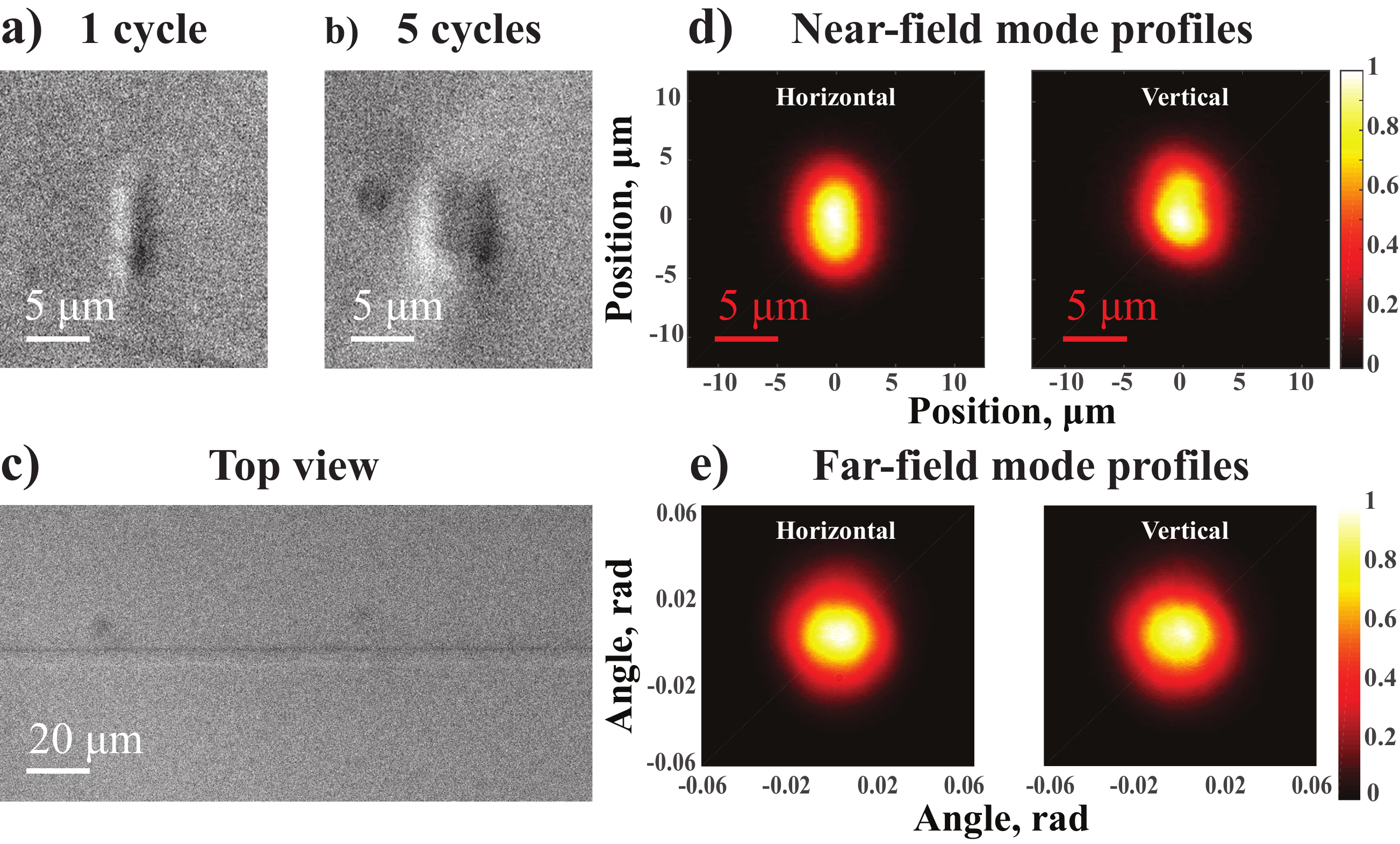}
\caption{(a)-(c) Images of the fabricated structures taken in an optical transmission microscope equipped with a DIC module. (a) and (b) show cross-section geometry of a modified region after a single pass and five shifted passes, respectively. The top view of the fabricated waveguide is shown in (c). CCD images for the near-field mode profiles (d) for horizontally and vertically polarized modes (with respect to the structure profile, shown in (b)). (e) Imaged mode profiles in the far-field.\label{fig::cross_sections_and_profiles}}
\end{figure}

Our initial fabrication setup did not include any beam shaping unit and the Gaussian writing beam was matched with the focusing objective input aperture. In this scheme either uniform but very small modification was induced in a low pulse energy regime or, for larger pulse energies, self-focusing and filamentation nonlinear processes were launched, which were responsible for the formation of an elongated high-contrast core \cite{Sudrie2002} surrounded by a large low-contrast halo similar to the one reported in \cite{Eaton2008} for N-BK7 glass. We applied a slit beam shaping technique to effectively reduce the numerical aperture of the focusing objective in the direction transverse to the waveguide axis. Even though the fixed slit shaping is not applicable for printing waveguides with a large turning angle \cite{Salter2012} it is still the simplest and the most effective beam profile management tool, enabling continuous adjustment of the beam waist in a desired direction. Searching for the optimal single scan process implied sweeping through the slit width, pulse energy and translation speed parameter space, while keeping the desired waveguide printing depth under the surface of the chip constant. Waveguides were printed 175~$\mu$m below the surface of the sample (Thermo Scientific, ISO 8037/l). Optimal translation speed was determined by visual inspection of uniformity of the inscribed structures. Sample structures for different slit width, pulse energy and translation speed parameters are depicted in Fig.~\ref{fig::parameter_map}, where the pulse energy was measured after the slit. The optimal performance was observed for the following values of the parameters: slit width of 325~$\mu$m, pulse energy of 42~nJ, and translation speed of 1~mm/s. The optimal translation speed was determined as a trade-off value between minimal longitudinal shear in the multiscan regime and maximal refractive index change. Using these settings we were able to produce uniform structures with elliptical $2 \times 8$~$\mu \mathrm{m}^{2}$ cross-section for a single scan (Fig.~\ref{fig::cross_sections_and_profiles}a). To reduce the eccentricity of the waveguides we performed a multiscan procedure with 1~$\mu$m transverse shift of each consecutive track producing approximately $7 \times 8$ $\mu \mathrm{m}^{2}$ modified refractive index area (Fig.~\ref{fig::cross_sections_and_profiles}b). A 1~$\mu$m transverse shift ensured the required overlap between successive modified volumes. End faces of the samples were ground and optically polished after fabrication.

\begin{center}
\begin{table*}
	\caption{Summary of the waveguide fundamental mode measurement results}
	\label{tab::mode_measurements}
	\begin{tabular}{ l | c | c | c | c }
	\toprule
		Polarization & MFD$_{x}$, $\mu$m & MFD$_{y}$, $\mu$m & Coupling loss & $\Delta n$ \\
	\hline
		horizontal & 9.2$\pm$0.1 & 12.0$\pm$0.1 & 1.12 dB (12.0~$\mu\mathrm{m}$ MFD) & \multirow{2}{*}{$1\cdot 10^{-3}$} \\
	\cline{1-4}
		vertical & 8.8$\pm$0.1 & 12.4$\pm$0.1 & 1.10 dB (11.2~$\mu\mathrm{m}$ MFD) & \\
	\bottomrule
	\end{tabular}
\end{table*}
\end{center}

\section{Waveguide characterization \label{sec::waveguide_char}}

\subsection{Waveguide fundamental mode profile \label{subsec::mode_profile}}

Quality of the fabricated waveguides was characterized with a setup shown in Fig.~\ref{fig::experimental_setup}(c). The output of a 808 nm diode laser was spatially filtered with a single-mode optical fiber and coupled to a waveguide with a 0.1 NA objective. The sample is mounted on a high-resolution 6-axis mechanical positioner (Luminos I6000). A Glan prism ensures the linear polarization of coupled laser light. Another objective lens with 0.25 NA images the near-field mode profiles on a CCD camera (Fig.~\ref{fig::cross_sections_and_profiles}(d)) with a 36X magnification factor. Far-field profiles (Fig.~\ref{fig::cross_sections_and_profiles}(e)) are captured by the CCD camera after collimating the waveguide output by a 10X objective with f = 18~mm focal length and are used to confirm the single-mode propagation regime. 

The measured mode field diameter (MFD) in the near-field $w_{wg}$ allows for a rough estimation of the refractive index contrast $\Delta n$ induced in the core of the waveguide: the numerical aperture of the waveguide is $NA_{wg} = \sqrt{2n\Delta n} = \frac{2\lambda}{\pi w_{wg}}$, where $n$ is the refractive index of the bulk material ($n=1.517$ for soda-lime glass at the wavelength of interest) and $\lambda$ is the wavelength. MFD measurement results and inferred $\Delta n$ values are summarized in Table~\ref{tab::mode_measurements}. 

Intrinsic coupling loss arising due to the mode mismatch between the coupled source and the waveguide is estimated by numerically evaluating the mode overlap coefficient $\eta_{c}$ of a measured near-field mode profile and the single-mode optical fiber mode profile at 808~nm:
\begin{equation}\label{eq::coupling_loss}
\eta_{c} = \frac{\int_{-\infty}^{\infty}\sqrt{I_{f}(x,y)}\sqrt{I_{wg}(x,y)}dxdy}{\sqrt{\int_{-\infty}^{\infty}I_{f}\left(x,y\right)dxdy}\sqrt{\int_{-\infty}^{\infty}I_{wg}\left(x,y\right)dxdy}},
\end{equation}
where $I_{f}(x,y)=\exp\left(-\frac{2\left(x+y\right)^{2}}{w_{f}^{2}}\right)$ is the normalized fundamental mode intensity of a single-mode fiber (a typical value of $w_{f}=2.5$~$\mu$m was taken for a standard Thorlabs SM780HP fiber) and $I_{wg}$ is the normalized measured near-field mode profile of the fabricated waveguide. Peak-to-peak coupling yielded the lowest possible coupling loss coefficient value 2.3 dB for the vertical mode and 2.4 dB for the horizontal mode.

If a free-space coupling scheme is employed, the coupling loss may be minimized by appropriate mode-match{\-}ing. In this case the loss may be estimated by minimizing the value of $\eta_c$ given by (\ref{eq::coupling_loss}) with respect to the input mode waist $w_f$. The estimates of coupling loss for a perfectly matched Gaussian beam are provided in Table~\ref{tab::mode_measurements}. Similar result may be achieved for a waveguide coupled with a single-mode fiber using a specially designed GRIN lens pigtail.

\begin{figure}[b]
\includegraphics[width=0.9\columnwidth]{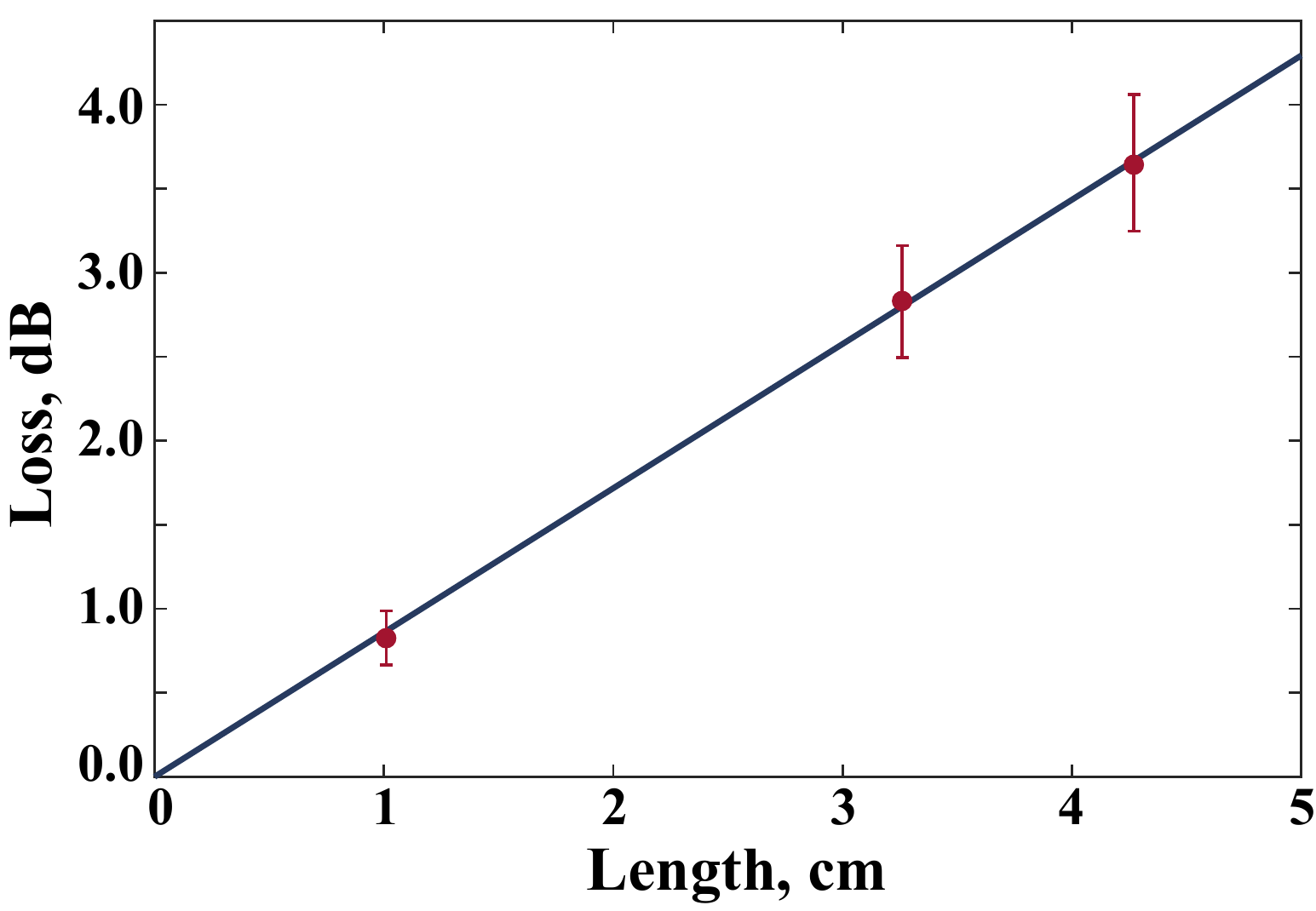}
\caption{Propagation loss measurement results. Error bars are obtained by statistical averaging over 10 waveguides fabricated with the same parameters set. The error bar value comprises both waveguide face polishing defects and relative inner discrepancy between different printed structures.\label{fig::propagation_loss}}
\end{figure}

\begin{center}
\begin{table*}[t]
	\caption{Comparative analysis of waveguide loss characteristics.}
	\label{tab::result_comparison}
	\begin{tabular}{ m{3cm} | >{\centering}m{3cm} | >{\centering}m{4cm} | >{\centering}m{5cm} @{}m{0pt}@{} }
	\toprule
		{} & Coupling loss & Propagation loss & Bending loss & \\
	\hline
		This work & 1.10~dB & 0.86$\pm$0.03~dB/cm (808 nm) & 2.7~dB/cm (30~mm radius)& \\[10pt]
	\hline
		Best reported (soda-lime glass)  & --- & 1.64~dB/cm (800 nm)\cite{Kowalevicz2005} & $\approx$ 3~dB/cm (30~mm radius) \cite{Tong2006}& \\[20pt]
	\hline 
		State-of-the-art (all materials, single-mode at 800~nm) & 0.7 dB (Thorlabs SM800-5.6-125 fiber)\cite{Sansoni2010} & 0.5~dB/cm \cite{Sansoni2010} & 0.3~dB/cm (30~mm radius)\cite{Sansoni2010}& \\[10pt]
	\bottomrule
	\end{tabular}
\end{table*}
\end{center}
 
\subsection{Propagation loss \label{subsec::prop_loss}}
To estimate the propagation loss $\alpha$ in straight waveguides we used a cut-back method, assuming that prepared structures are identical and exhibit equal coupling efficiency. However, in practice, the fabrication and post-processing conditions may vary slightly and affect the coupling efficiency and the propagation loss. Thus, to perform a relevant estimation of $\alpha$ we printed three chips of different lengths, each containing ten waveguides. Averaging over a series of measurements we evaluated the propagation loss taking into account possible fabrication imperfections. Fig.~\ref{fig::propagation_loss} shows the measurement result. Linear fit of the experimental data gives $\alpha = 0.86\pm 0.03$~dB/cm. Good agreement with a linear fitting model indicates that all waveguides exhibit identical average propagation loss, whereas our polishing procedure leaves some residual roughness on the input face of each chip, imposing a constant coupling efficiency uncertainty. The relative uncertainty gradually reduces with increasing length of the chip, indicating the presence of a minor distinction between the different waveguides.

\begin{figure}[htbp]
\includegraphics[width=0.9\columnwidth]{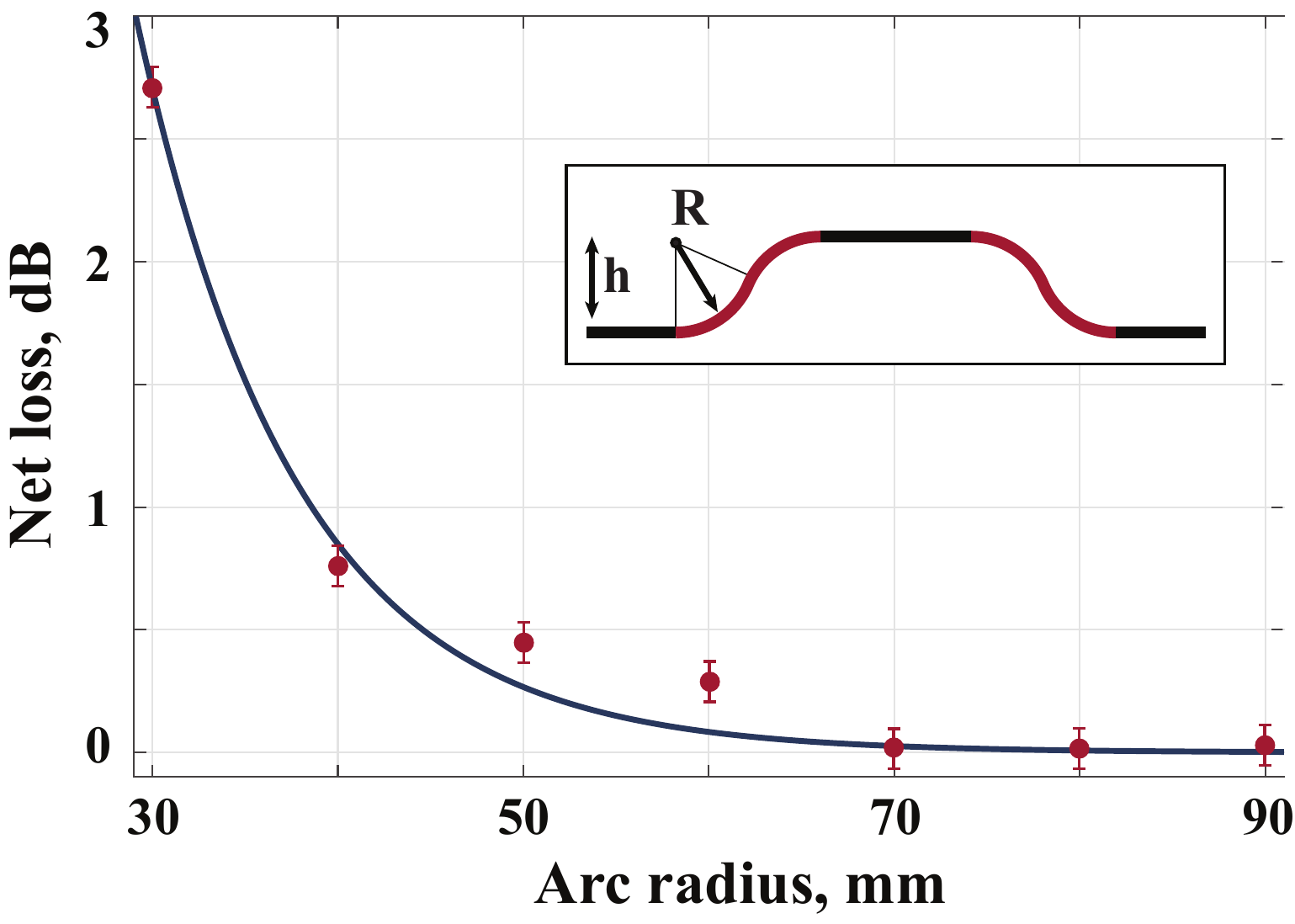}
\caption{Net bending loss measurement results. Inset illustrates the geometry of curved waveguides. Red segments consist of two circular arcs of radius $R$ and straight segments are separated by $h$.\label{fig::bend_loss}}
\end{figure}

\subsection{Bending loss \label{subsec::bend_loss}}
Even primitive integrated photonic devices, such as directional couplers and beam-splitters, require curved wave{\-}guides. Bending imposes excess loss due to light propagation in curved sections and mode mismatch effects in joint interfaces. To experimentally quantify bending loss we printed a series of curved waveguides. The inset in Fig.~\ref{fig::bend_loss} illustrates the geometry of the prepared structures, representing a half of a conventional directional coupler. An array of structures with varying parameters were produced. Each structure in the array consisted of three straight sections and four circular arcs. The radius of arc curvature varied from 30 mm to 90 mm. We measured the overall transmission for each structure. To derive the loss associated solely with bending, we subtract the loss accrued in the straight parts, which were calculated based on the previously obtained results, from the measured overall loss. The resulting bending loss level  is shown in Fig.~\ref{fig::bend_loss} as a function of radius of curvature. 

For applications without stringent requirements to minimize photonic circuit footprint, a 70 mm curvature radius should be used due to negligible bending loss ($\approx 0.02$ dB additional net loss). The arc radius may be reduced down to 50 mm imposing additional $\approx 0.4$ dB loss if the least possible circuit area is considered.

It is instructive to compare the experimental dependence with an analytical theory for the low-index-contrast case (see, for example, \cite{Snyder_Love}). Provided that the mode mismatch in the junction between straight and bent waveguides is negligible, the radiation loss coefficient is given by 
\begin{equation}\label{eq::bending_loss}
	\alpha = \frac{A}{\sqrt{R}}\exp\left(-BR\right),
\end{equation}
where $R$ is the radius of curvature, parameter $A$ depends on the waveguide cross-section geometry, and 
\begin{equation}\label{eq::B_parameter}
	B = \frac{2k\left(n_{eff}^2 - n^2\right)^{3/2}}{3n^2},
\end{equation}
where $k=2\pi/\lambda$ is the wavenumber, $n$ is the refractive index of the bulk, and $n_{eff}$ -- the effective mode index of the waveguide. From simple geometrical considerations the total length of the curved part of the waveguide may be calculated as $L_{bend} = 4R\arccos(1-h/2R)\approx4\sqrt{hR}$, thereby the losses read:
\begin{equation}\label{eq::bending_loss_dB}
	\delta\left[\mathrm{dB}\right] = 10\log_{10}\left(e\alpha L_{bend}\right)\approx 17.37\sqrt{h}A\exp(-BR).
\end{equation}
We fit the bend loss data (Fig.~\ref{fig::bend_loss}) by the exponential dependence (\ref{eq::bending_loss_dB}), which in combination with (\ref{eq::B_parameter}) yields $n_{eff}-n = (4.3\pm0.8)\cdot10^{-4}$.

As a consistency check we compare the obtained value of $n_{eff}$ with the one 	derived from the numerical eigenmode simulation for a waveguide having a rectangular cross-section of $7\times8$~$\mu\mathrm{m}^2$ and a refractive index contrast taken from Table~\ref{tab::mode_measurements}. The simulation gives $n_{eff}-n = 4.1\cdot10^{-4}$, which is in good agreement with the value obtained from the bending loss analysis.

\section{Conclusion}
We presented the fabrication process of low-loss single-mode optical waveguides in extra-white soda-lime glass. The waveguide cross-section is shaped to the desired form exploiting a multiscan writing process. With multiscan waveguides we were able to achieve well-defined symmetric mode shapes and reduced propagation and bending loss. We took a detailed account of all types of loss in the waveguides. Importantly, the well defined rectangular shape of the produced waveguides allowed us to quantitatively compare the measured values of bending loss with a simple analytical theory, as well as numerical calculations, and achieve an excellent correspondence.   

It would be interesting to compare the properties of our multiscan waveguides with previously reported data for waveguides in soda-lime glass and other optical materials. Such a comparison for an operating wavelength of 800~nm is presented in Table~\ref{tab::result_comparison}. Although the state-of-the-art values obtained with other materials (borosilicate glass) are somewhat better, we have achieved a significant improvement of the waveguides quality in soda-lime glass. Our results may improve the current femtosecond waveguide writing technology extending it to a low-cost material segment.

Authors are grateful to K.~S.~Kravtsov for enlightening discussions and careful reading of the manuscript, and to N.~S.~Chupriyanov for considerable help at the early stages of the experiment. 
This work was supported by the Russian Science Foundation (project 16-12-00017).

\bibliographystyle{phaip}
\bibliography{wm_bibliography}

\begin{thebibliography}{10}

\bibitem{Doerr2015}
C.~R. Doerr,
\newblock Frontiers in Physics {\bf 3}, 1 (2015).

\bibitem{Washburn2011}
A.~L. Washburn and R.~C. Bailey,
\newblock The Analyst {\bf 136}, 227 (2011).

\bibitem{OBrien2013}
J.~O'Brien, B.~Patton, M.~Sasaki, and J.~Vuckovic,
\newblock New Journal of Physics {\bf 15}, 15 (2013).

\bibitem{DellaValle2008}
G.~{Della Valle}, R.~Osellame, and P.~Laporta,
\newblock Journal of Optics A: Pure and Applied Optics {\bf 11}, 013001 (2008).

\bibitem{Shah2005}
L.~Shah, A.~Arai, S.~Eaton, and P.~Herman,
\newblock Optics express {\bf 13}, 1999 (2005).

\bibitem{Eaton2006}
S.~M. Eaton, W.-J. Chen, H.~Zhang, and P.~R. Herman,
\newblock Conference on Lasers and Electro-Optics/Quantum Electronics and Laser
  Science Conference and Photonic Applications Systems Technologies {\bf 18},
  JTuD8 (2006).

\bibitem{Chan2003}
J.~W. Chan, T.~R. Huser, S.~H. Risbud, J.~S. Hayden, and D.~M. Krol,
\newblock Applied Physics Letters {\bf 82}, 2371 (2003).

\bibitem{Efimov2001}
O.~M. Efimov et~al.,
\newblock Optical Materials {\bf 17}, 379 (2001).

\bibitem{Schaffer2003}
C.~B. Schaffer, J.~F. Garc{\'{i}}a, and E.~Mazur,
\newblock Applied Physics A: Materials Science and Processing {\bf 76}, 351
  (2003).

\bibitem{Taccheo2004}
S.~Taccheo et~al.,
\newblock Optics Letters {\bf 29}, 2626 (2004).

\bibitem{Psaila2007}
N.~D. Psaila et~al.,
\newblock Conference on Lasers and Electro-Optics, 2007, CLEO 2007 {\bf 14},
  1515 (2007).

\bibitem{Minoshima2001}
K.~Minoshima, a.~M. Kowalevicz, I.~Hartl, E.~P. Ippen, and J.~G. Fujimoto,
\newblock Optics letters {\bf 26}, 1516 (2001).

\bibitem{Kowalevicz2005}
A.~M. Kowalevicz, V.~Sharma, E.~P. Ippen, J.~G. Fujimoto, and K.~Minoshima,
\newblock Opt. Lett. {\bf 30}, 1060 (2005).

\bibitem{Tong2006}
L.~Tong, R.~R. Gattass, I.~Maxwell, J.~B. Ashcom, and E.~Mazur,
\newblock Optics Communications {\bf 259}, 626 (2006).

\bibitem{Lazcano2016}
H.~E. Lazcano and G.~V. V{\'{a}}zquez,
\newblock Appl. Opt. {\bf 55}, 3268 (2016).

\bibitem{Gattass2008}
R.~R. Gattass and E.~Mazur,
\newblock Nature Photonics {\bf 2}, 219 (2008).

\bibitem{Osellame2003}
R.~Osellame et~al.,
\newblock Journal of the Optical Society of America B {\bf 20}, 1559 (2003).

\bibitem{Ams2008}
M.~Ams, G.~D. Marshall, P.~Dekker, and M.~J. Withford,
\newblock PIERS Online {\bf 4}, 146 (2008).

\bibitem{Salter2012}
P.~S. Salter et~al.,
\newblock Optics Letters {\bf 37}, 470 (2012).

\bibitem{simmonds2011}
R.~D. Simmonds, P.~S. Salter, A.~Jesacher, and M.~J. Booth,
\newblock Optics Express {\bf 19}, 24122 (2011).

\bibitem{Okhrimchuk2012}
A.~Okhrimchuk, V.~Mezentsev, A.~Shestakov, and I.~Bennion,
\newblock Optics express {\bf 20}, 3832 (2012).

\bibitem{Nasu2005}
Y.~Nasu, M.~Kohtoku, and Y.~Hibino,
\newblock Opt. Lett. {\bf 30}, 723 (2005).

\bibitem{Bookey2007}
H.~T. Bookey, R.~R. Thomson, N.~D. Psaila, and A.~K. Kar,
\newblock {Multi-scan femtosecond laser waveguide inscription in z- cut Lithium
  Niobate},
\newblock in {\em CLEO/Pacific Rim}, 2007.

\bibitem{Brown2012}
G.~Brown, R.~R. Thomson, A.~K. Kar, N.~D. Psaila, and H.~T. Bookey,
\newblock Optics Letters {\bf 37}, 491 (2012).

\bibitem{Douglass2015}
G.~Douglass, F.~Dreisow, S.~Gross, S.~Nolte, and M.~J. Withford,
\newblock Optics Express {\bf 23}, 21392 (2015).

\bibitem{Eaton2008}
S.~M. Eaton et~al.,
\newblock Appl. Opt. {\bf 47}, 2098 (2008).

\bibitem{McMillen2012}
B.~McMillen, B.~Zhang, K.~P. Chen, A.~Benayas, and D.~Jaque,
\newblock Opt. Lett. {\bf 37}, 1418 (2012).

\bibitem{Crespi2013}
A.~Crespi et~al.,
\newblock Nature Photonics {\bf 7}, 545 (2013).

\bibitem{Sudrie2002}
L.~Sudrie et~al.,
\newblock Phys. Rev. Lett. {\bf 89}, 186601 (2002).

\bibitem{Sansoni2010}
L.~Sansoni et~al.,
\newblock Physical Review Letters {\bf 105}, 1 (2010).

\bibitem{Snyder_Love}
A.~Snyder and J.~Love,
\newblock {\em Optical waveguide theory},
\newblock Chapman \& Hall, 1983.

\end{thebibliography}

\end{document}